\documentclass[usenatbib]{mnras}

\usepackage{mathptmx}
\usepackage[T1]{fontenc}
\usepackage[british]{babel}

\usepackage[utf8]{inputenc}
\usepackage{lmodern}

\usepackage{ae,aecompl}

\usepackage{graphicx}	
\usepackage{amsmath}	
\usepackage{amssymb}	

\usepackage{epstopdf}

\title[TTVs, RVs \& long-term stability of Kepler-410]{Transit timing variations, radial velocities and long-term dynamical stability of the system Kepler-410\thanks{This article is based on the data collected with Perek 2-m telescope.}}

\author[Gajdo\v{s} et al.]{Pavol Gajdo\v{s}$^{1}$\thanks{E-mail: pavol.gajdos@student.upjs.sk},
Martin Va\v{n}ko$^{2}$,
Theodor Pribulla$^{2}$,
Daniel Dupkala$^{3, 4}$,
J\'{a}n \v{S}ubjak$^{3, 4}$,\newauthor
Marek Skarka$^{3, 5}$,
Petr Kab\'{a}th$^{3}$,
\v{L}ubom\'{i}r Hamb\'{a}lek$^{2}$,
\v{S}tefan Parimucha$^{1}$
\\
$^{1}$Institute of Physics, Faculty of Science, Pavol Jozef \v{S}af\'arik University, Ko\v{s}ice, Slovakia\\
$^{2}$Astronomical Institute, Slovak Academy of Sciences, 059 60 Tatransk\'a Lomnica, Slovakia\\ 
$^{3}$Astronomical Institute, Czech Academy of Sciences, Fri\v{c}ova 298, 25165, Ond\v{r}ejov, Czech Republic\\         
$^{4}$Astronomical Institute of Charles University, V Hole\v{s}ovi\v{c}k\'{a}ch 2, 180 00, Praha, Czech Republic\\
$^{5}$Department of Theoretical Physics and Astrophysics, Masaryk Univerzity, Kotl\'{a}\v{r}sk\'{a} 2, 61137 Brno, Czech Republic              
}

\date{Accepted 2019 January 22. Received 2019 January 21; in original form 2018 December 14.}

\pubyear{2019}

\begin{document}
\setcounter{page}{4352}
\label{firstpage}
\volume{484}
\pagerange{\pageref{firstpage}--\pageref{lastpage}}
\maketitle

\begin{abstract}  
Transit timing variations of Kepler-410Ab were already reported in a few papers. Their semi-amplitude is about 14.5 minutes. In our previous paper, we found that the transit timing variations could be caused by the presence of a stellar companion in this system. Our main motivation for this paper was to investigate variation in a radial-velocity curve generated by this additional star in the system. We performed spectroscopic observation of Kepler-410 using three telescopes in Slovakia and Czech Republic. Using the cross-correlation function, we measured the radial velocities of the star Kepler-410A. We did not observe any periodic variation in a radial-velocity curve. Therefore, we rejected our previous hypothesis about additional stellar companion in the Kepler-410 system. We ran different numerical simulations to study mean-motion resonances with Kepler-410Ab. Observed transit timing variations could be also explained by the presence of a small planet near to mean-motion resonance 2:3 with Kepler-410Ab. This resonance is stable on a long-time scale. We also looked for stable regions in the Kepler-410 system where another planet could exist for a long time.
\end{abstract}


\section{Introduction}

The transiting exoplanet Kepler-410Ab, a Neptune-sized planet on the 17.8336 days orbit, was discovered in 2013 by \textit{Kepler} telescope and confirmed by \citet{Eylen2014}. The host star Kepler-410A (KIC 8866102, KOI-42, HD 175289) is a young \mbox{$2.76 \pm 0.54$~Gyr} old star with radius $1.352 \pm 0.010$~R$_{\sun}$, mass $1.214 \pm 0.033$~M$_{\sun}$, a spectral type F6IV and \textit{V} magnitude 9.5 \citep{Molenda2013}. The second GAIA Data Release \citep{Gaia2018} gives a distance of $148.16\pm0.49$~pc.

Using adaptive optics, \citet{Adams2012} distinguished a low-mass stellar companion Kepler-410B separated by an angular distance of 1.63\,arcsec. \cite{Eylen2014} found that this star is probably a red dwarf and ruled it out as a host star for the exoplanet.

We analysed transit timing variations (TTVs) in this system in our previous paper \citep{Gajdos2017}. We found periodic variations with semi-amplitude of about 14.5 minutes which we explained using a model of light-time effect (LiTE) \citep{Irwin1952} and perturbing model given by \cite{Agol2005}, by the presence of another stellar-mass (1~--~2~M$_{\sun}$) object in Kepler-410 system.

The aim of this paper is to observe variations in a radial-velocity curve caused by this stellar companion and to analyse them. A second motivation of the present study was to perform simulations leading to the generation of stability maps for the system. We also studied the possibility of mean-motion resonances with Kepler-410Ab and a stability of them.

In Section \ref{ttv}, we summarize the results of previous studies about TTVs of Kepler-410Ab. In Section \ref{rv} we present our spectroscopic observations and radial velocities of the system. We analyse resonant perturbation in Section \ref{res}. Section \ref{stab} is concerned with the long-term stability of the system. Our results are discussed in Section \ref{concl}.

\section{Transit timing variations}
\label{ttv}

TTVs of Kepler-410Ab were for the first time reported by \citet{Mazeh2013}. They studied the TTVs using a sinusoidal model and found a TTV semi-amplitude of $13.91 \pm 0.91$ minutes and a period of about 960 days. In a subsequent analysis, \citet{Eylen2014} used a 'zigzag' model. They obtained a slightly larger semi-amplitude of 16.5 minutes, and a period of 957 days. They also noted that the shape of the TTVs in not sinusoidal, which could be caused by an eccentric orbit of Kepler-410Ab. \cite{Holczer2016} found out TTVs with semi-amplitude of $15.59 \pm 0.49$ minutes and a period of $960.6 \pm 9.2$ days.

\begin{figure}
\includegraphics[width=\columnwidth]{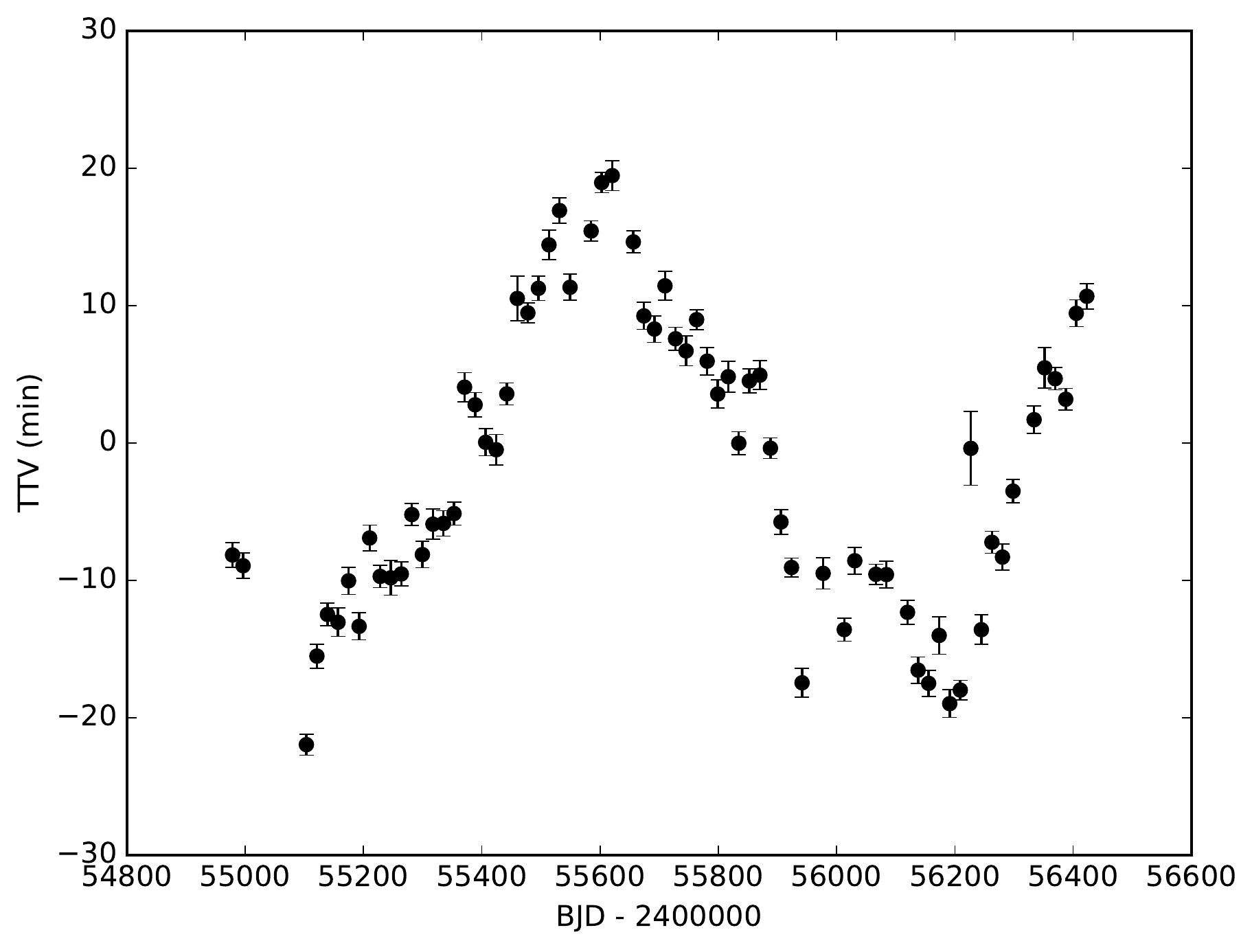}
\caption{Transit timing variations of Kepler-410Ab presented in our previous paper \citep{Gajdos2017}.}
\label{fig:ttv}
\end{figure}

In our previous study \citep{Gajdos2017}, we found periodic variations with semi-amplitude of about 14.5 minutes and period of about 972 days (see Figure \ref{fig:ttv}). We used two different models to explain the observed variations. The first one was a model of light-time effect \citep{Irwin1952}. Using the LiTE solution, we arrived at the mass function $f(M_3)=0.879\pm0.030$~M$_{\sun}$. For a coplanar orbit, we got the minimal mass of the body $M_3=2.151\pm0.078$~M$_{\sun}$. If we consider a main-sequence star, this would correspond to the spectral type A4. Star with such luminosity was not detected in spectra obtained by \citet{Molenda2013} nor in our new spectra (see Section \ref{rv}). Moreover, no signs of eclipses with $\sim$~970-day period were found during the $\sim$1460-day long observation by \textit{Kepler}. Thus, we concluded that the perturbing body could be a non-eclipsing binary star. Kepler-410 is a quadruple star with one planet (orbiting Kepler-410A) in this scenario.

We also used an analytical approximation of the perturbation model given by \citet[their eq. 25]{Agol2005}. This method was originally developed for the analysis of multiple planetary systems but is not constrained to any limit of mass of the perturbing body. We assumed coplanar orbits and calculated the mass of the third body $M_3=0.906\pm0.155$~M$_{\sun}$. If this model is correct, Kepler-410 is a triple star system with one planet.

\section{Spectroscopic observations and radial velocities}
\label{rv}

\begin{table}
\caption{Radial velocity measurements from observatories: Ond. - Ond\v{r}ejov Observatory, SL - Star\'{a} Lesn\'{a} Observatory, SP - Skalnat\'{e} Pleso Observatory}
\label{tab:rv}
\begin{center}
{\footnotesize
\begin{tabular}{ccccc}
	\hline
	HJD-2400000 & RV (km/s) & Error (km/s) & SNR & Obs. \\ \hline
	57561.4897  &  -40.40   &     0.37     & 33  &  SL  \\
	57563.4849  &  -40.17   &     0.37     & 28  &  SL  \\
	57563.5207  &  -40.52   &     0.37     & 22  &  SL  \\
	57575.3651  &  -40.68   &     0.35     & 22  &  SL  \\
	57608.4225  &  -41.04   &     0.37     & 29  &  SL  \\
	57609.3982  &  -40.70   &     0.37     & 35  &  SL  \\
	57624.3387  &  -39.90   &     0.34     & 21  &  SL  \\
	57625.3144  &  -40.33   &     0.36     & 23  &  SL  \\
	57626.3263  &  -40.63   &     0.35     & 34  &  SL  \\
	57641.3165  &  -40.47   &     0.37     & 34  &  SL  \\
	57643.3638  &  -40.92   &     0.36     & 37  &  SL  \\
	57657.2555  &  -40.98   &     0.35     & 28  &  SL  \\
	57874.4172  &  -38.60   &     0.48     & 17  & Ond  \\
	57884.4760  &  -40.78   &     0.28     & 34  & Ond  \\
	57888.5807  &  -40.36   &     0.40     & 18  &  SL  \\
	57901.5031  &  -41.40   &     0.38     & 18  &  SL  \\
	57906.4804  &  -39.11   &     0.36     & 14  &  SL  \\
	57926.3788  &  -40.10   &     0.21     & 27  & Ond  \\
	57935.4804  &  -40.80   &     0.23     & 24  & Ond  \\
	57950.3847  &  -40.33   &     0.30     & 15  & Ond  \\
	57952.4047  &  -40.20   &     0.21     & 50  & Ond  \\
	57967.4011  &  -40.63   &     0.41     & 23  &  SL  \\
	57968.3506  &  -38.67   &     0.34     & 20  &  SL  \\
	57969.4190  &  -40.30   &     0.24     & 16  & Ond  \\
	57971.4031  &  -39.14   &     0.38     & 18  &  SL  \\
	57983.3749  &  -38.89   &     0.19     & 38  & Ond  \\
	57988.3695  &  -39.60   &     0.26     & 12  & Ond  \\
	57989.3711  &  -40.24   &     0.36     & 32  &  SL  \\
	57995.3298  &  -40.76   &     0.19     & 44  & Ond  \\
	58015.2968  &  -38.10   &     0.35     & 33  & Ond  \\
	58017.2932  &  -39.36   &     0.34     & 21  & Ond  \\
	58036.2559  &  -41.19   &     0.35     & 24  &  SL  \\
	58202.6574  &  -40.47   &     0.37     & 25  &  SL  \\
	58217.6144  &  -40.72   &     0.34     & 31  &  SL  \\
	58229.5715  &  -40.55   &     0.35     & 39  &  SL  \\
	58245.5763  &  -39.92   &     0.38     & 37  &  SL  \\
	58249.5893  &  -40.01   &     0.41     & 22  &  SL  \\
	58252.5602  &  -40.69   &     0.36     & 26  &  SL  \\
	58319.4683  &  -40.17   &     0.21     & 37  & Ond  \\
	58320.4124  &  -40.77   &     0.21     & 44  & Ond  \\
	58326.4110  &  -39.53   &     0.23     & 33  & Ond  \\
	58334.4223  &  -38.96   &     0.24     & 37  & Ond  \\
	58343.3805  &  -40.56   &     0.38     & 33  &  SL  \\
	58347.3279  &  -41.67   &     0.18     & 39  &  SP  \\
	58347.3360  &  -40.34   &     0.39     & 26  &  SL  \\
	58347.4378  &  -39.17   &     0.21     & 30  & Ond  \\
	58351.3182  &  -40.34   &     0.20     & 34  & Ond  \\
	58352.3063  &  -40.40   &     0.21     & 34  & Ond  \\
	58357.3184  &  -40.58   &     0.24     & 22  & Ond  \\
	58358.3320  &  -40.24   &     0.38     & 26  &  SL  \\
	58361.3259  &  -42.14   &     0.20     & 17  &  SP  \\
	58361.3517  &  -40.87   &     0.37     & 30  &  SL  \\
	58366.3023  &  -41.33   &     0.27     & 33  & Ond  \\
	58374.2994  &  -40.40   &     0.38     & 32  &  SL  \\
	58379.3391  &  -40.00   &     0.38     & 35  &  SL  \\
	58380.3073  &  -40.95   &     0.38     & 32  &  SL  \\
	58388.4004  &  -39.12   &     0.24     & 47  & Ond  \\
	58392.2790  &  -40.09   &     0.38     & 29  &  SL  \\
	58403.2579  &  -42.13   &     0.19     & 48  &  SP  \\
	58427.2124  &  -40.39   &     0.18     & 34  &  SP  \\ \hline
\end{tabular}}
\end{center}
\end{table}

TTVs models proposed in our paper predict radial-velocity (RV) changes with semi-amplitude $\sim 25 - 30$~km/s \citep{Irwin1952-rv}. The period of the variation would be the same as the period of possible stellar companion (i.e. $\sim 972$ days). The precision at the level of 1~km/s is more than sufficient for detecting this variation. 

Our spectroscopic observations of Kepler-410 started in 2016 with a 60-cm telescope at Star\'{a} Lesn\'{a} Observatory. During years 2016 -- 2018 we made overall 60 RV measurements (listed in Table \ref{tab:rv}) using three telescopes.

We obtained 34 spectra at Star\'{a} Lesn\'{a} Observatory. The Star\'{a} Lesn\'{a} Observatory (20\degr 17\arcmin 21\arcsec E, 49\degr 09\arcmin 06\arcsec N, 810~meters a.s.l.) is part of the Astronomical Institute of the Slovak Academy of Sciences. It is equipped with 60-cm (f/12.5) Zeiss reflecting telescope with fiber-fed échelle spectrograph eShel \citep{Pribulla2015}. The spectra consisting of 24 orders cover the wavelength range from 4150 to 7600~\AA{}. The resolving power of the spectrograph is \mbox{$R = 10000 - 12000$}. The ThAr calibration unit provides about 100~m/s radial-velocity accuracy. Reduction pipeline using IRAF \citep{Tody1993} has been described by \cite{Pribulla2015}.

Kepler-410 system was also observed on 22 different nights during 2017 till 2018 period  in Ond\v{r}ejov Observatory (14\degr 46\arcmin 52\arcsec E, 49\degr 54\arcmin 55\arcsec N, 525 meters a.s.l.). We obtained spectra with OES spectrograph installed at 2-m Perek telescope. OES is an \'{e}chelle spectrograph with resolving power $R=44000$ sensitive between 3600~--~9500~\AA. OES is using ThAr spectra for wavelength calibration. The more detailed description is given in \cite{Kabath2018}. The spectra were reduced with IRAF.

We obtained also four spectra in 2018 with a 1.3m, f/8.36, Nasmyth-Cassegrain telescope equipped with a fiber-fed \'echelle spectrograph at Skalnat\'e Pleso Observatory (20\degr 14\arcmin 02\arcsec E, 49\degr 11\arcmin 22\arcsec N, 1786 meters a.s.l.). The spectrograph follows the MUSICOS design \citep{Baudrand1992}. The spectra were recorded by an Andor iKon 936 DHZ CCD camera, with a 2048$\times$2048 array, 15$\mu$m square pixels, 2.7e$^-$ read-out noise and gain close to unity. The spectral range of the instrument is 4250~--~7375~\AA ~(56 \'echelle orders) with the maximum resolution of $R = 38000$. Reduction pipeline is the same as in the case of spectra from Star\'{a} Lesn\'{a} Observatory.

\begin{figure}
\includegraphics[width=\columnwidth]{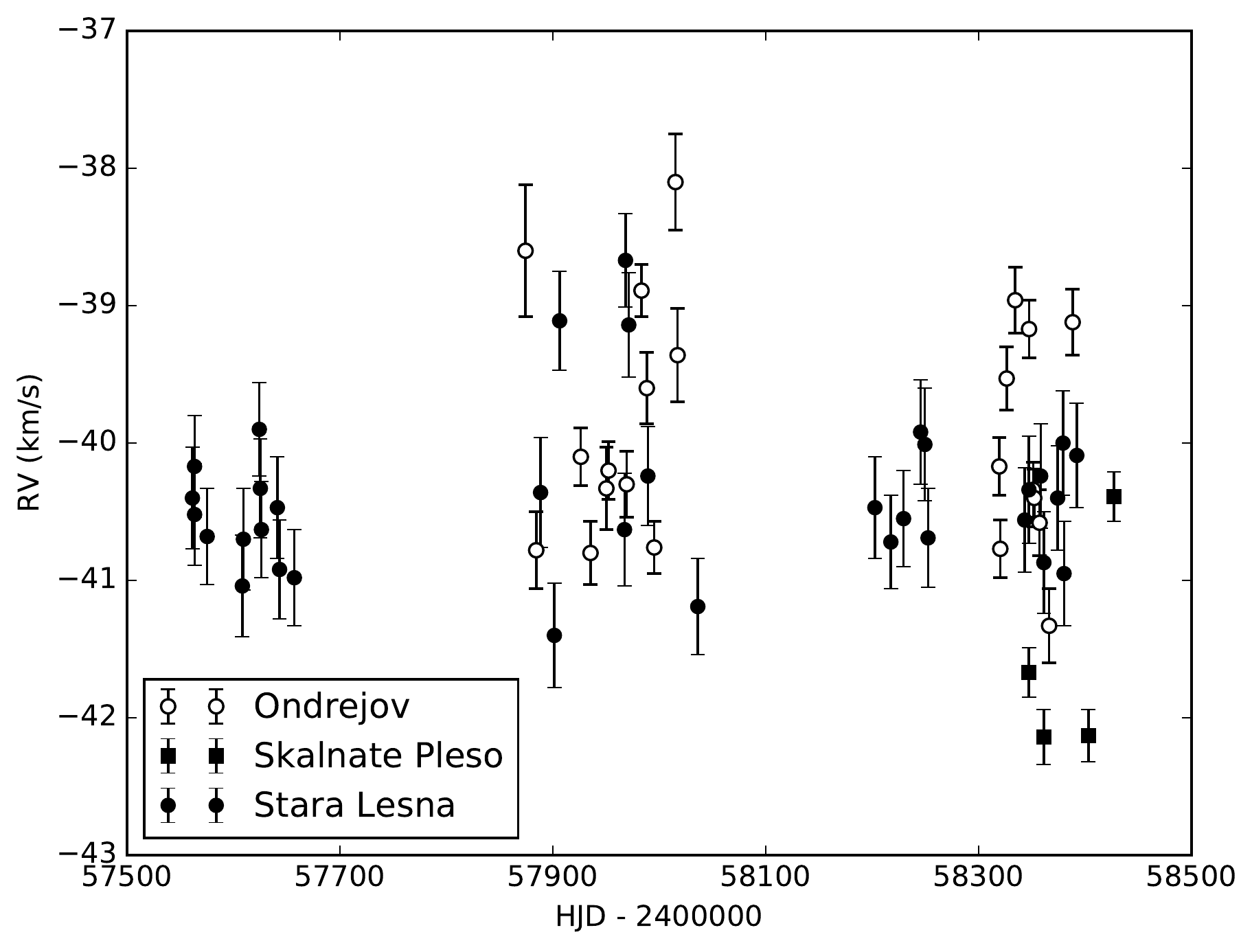}
\caption{Radial velocities of Kepler-410.}
\label{fig:rv}
\end{figure}

In all collected spectra, we could detect spectral lines only from one stellar source (Kepler-410A). A dwarf star Kepler-410B is too faint to be detected using our spectra with low SNR obtained by small telescopes. No spectroscopic signal from hypothetical star Kepler-410C (one star or close binary), proposed in our TTV analysis (see Section \ref{ttv}), was observed. It is the first observational evidence that TTVs are caused by a different mechanism.

We measured the radial velocity of Kepler-410 using cross-correlation function in iSpec \citep{Cuaresma2014}. We have selected only spectral region around magnesium triplet (5000~--~5500~\AA). As a template, we used a spectrum of $\sigma$ Boo (F4V spectral type star with V magnitude 4.47) observed at Skalnat\'e Pleso Observatory on 31 May 2018. Radial velocities were corrected according to heliocentric correction and radial velocity of $\sigma$ Boo \citep[370 m/s;][]{Nidever2002}. All our measurements are listed in Table \ref{tab:rv} and displayed on Figure \ref{fig:rv}. 

We did not observe any periodic variation on a radial-velocity curve with an amplitude greater than $\sim 600 - 800$~m/s over observing period. All obtained RVs are distributed close around their mean value $-40.3$~km/s. A few outlier points are caused by poor quality of these spectra. We also observed the bigger variance of the data obtained from Star\'{a} Lesn\'{a} observatory in 2017 comparing to the data from the same observatory in different years. It could be caused by some systematic error during the observations (e.g. problem with the thermal stability of spectrograph). The relatively big scatter of data collected on Ond\v{r}ejov observatory could be caused by the slightly different methodology of observations and spectra reduction than in both Slovak observatory. Possible reduction of these (mainly) systematic variations would decrease the maximal amplitude of hypothetical RV signal by about 100~m/s.

Multi-star explanation of observed TTVs, which would lead to RV with an amplitude at the level $\sim 25 - 30$ km/s, seems to be very unlikely. We can also exclude the existence of a brown dwarf (down to mass~$\approx 30$~M$_{Jup}$) on the nearly edge-on orbits (i.e. the inclination angle is 90\degr) with period up to 2 years and massive hot Jupiter in this system. The presence of hot Jupiter is of course excluded by the absence of its transits and by the transit timing of Kepler-410Ab (see Figure \ref{fig:limit}). Unfortunately, the precision of our RV measurements is not enough to put this mass constraint below that of given by TTVs analysis.

\section{Mean-motion resonances}
\label{res}

\begin{figure}
\includegraphics[width=\columnwidth]{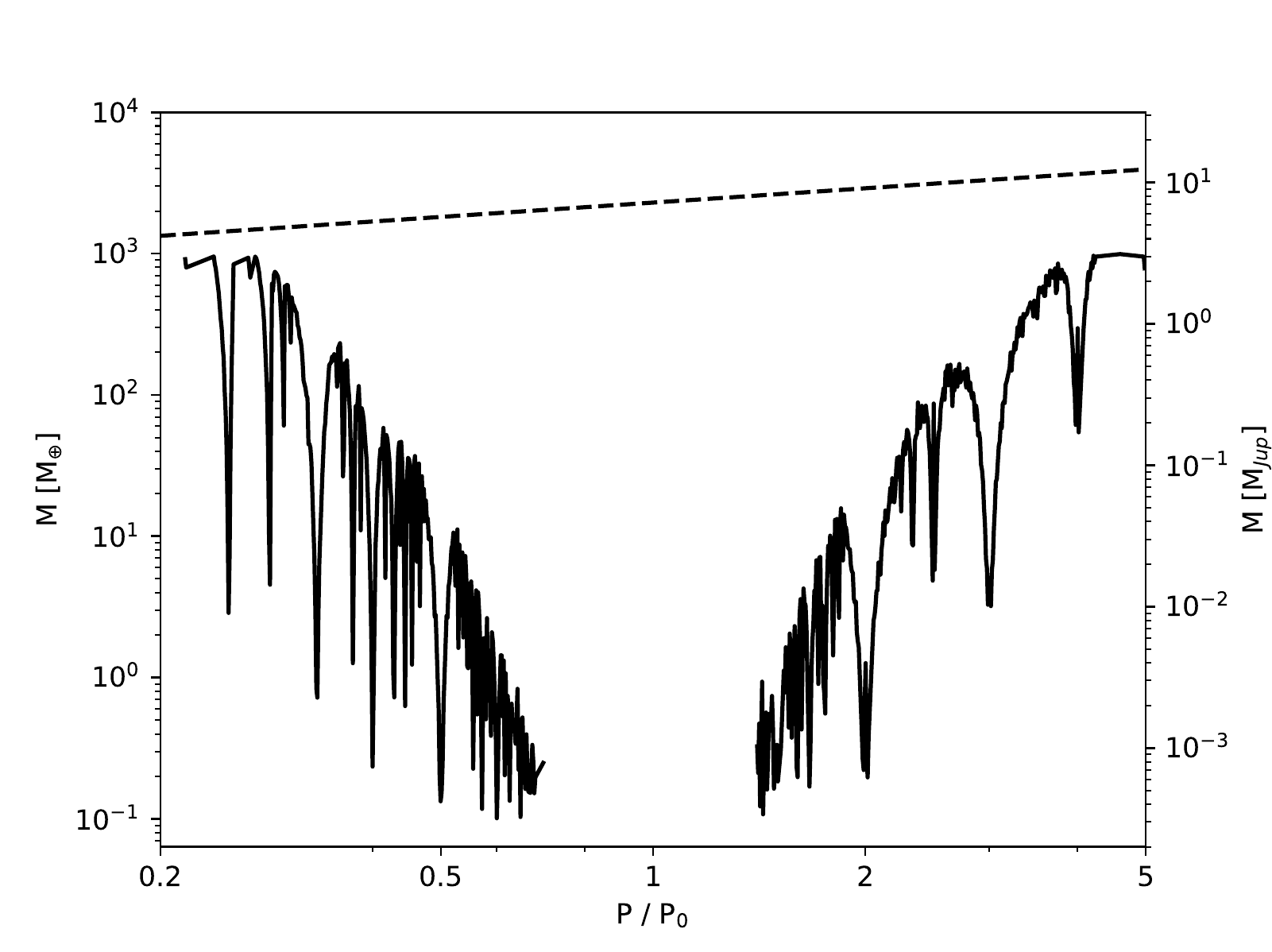}
\caption{The upper-mass limits of a hypothetical additional planet in the Kepler-410 system based on TTV analysis (solid curve) and RV measurements (dashed line).}
\label{fig:limit}
\end{figure}

\begin{figure}
\includegraphics[width=\columnwidth]{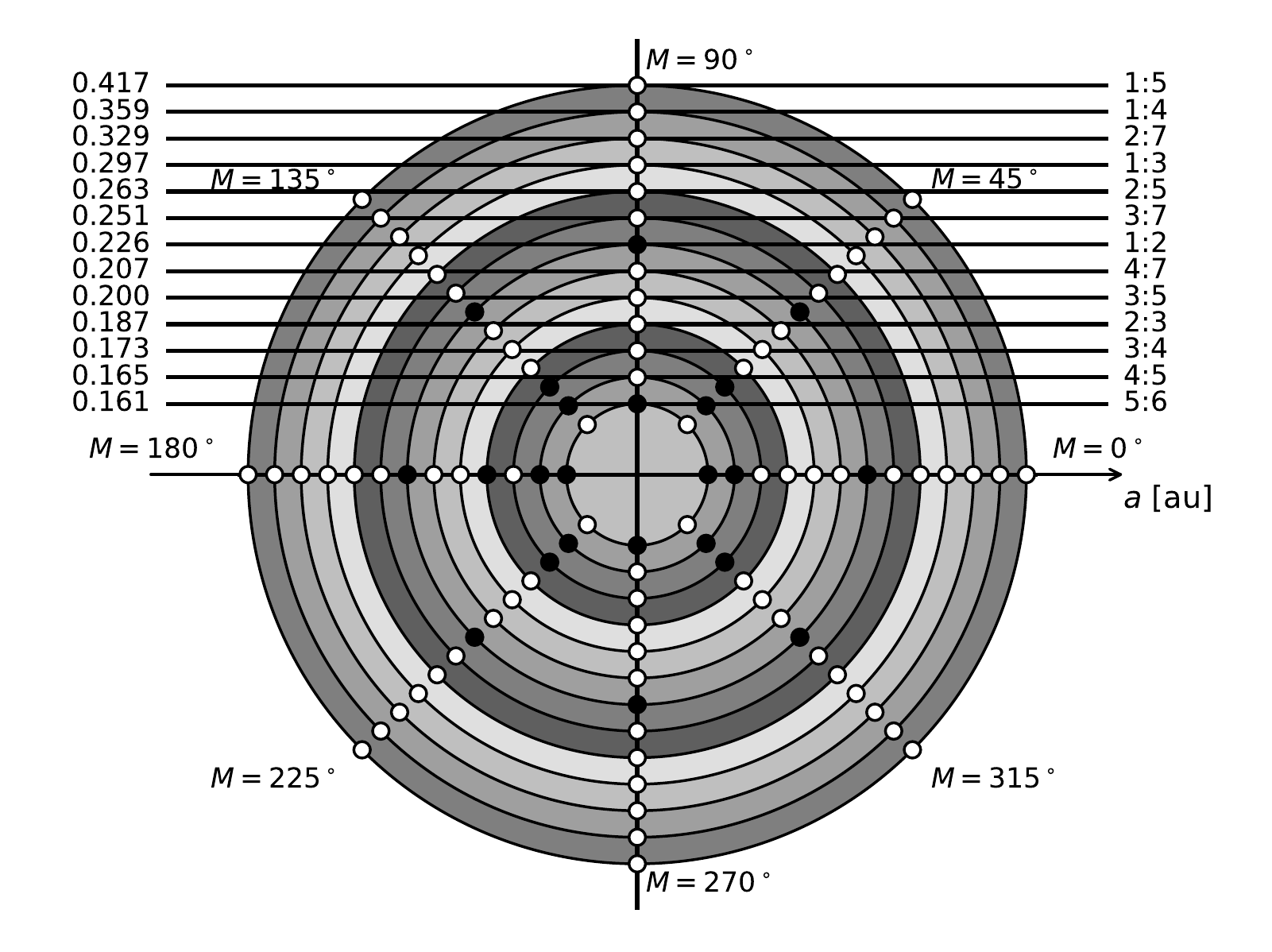}
\includegraphics[width=\columnwidth]{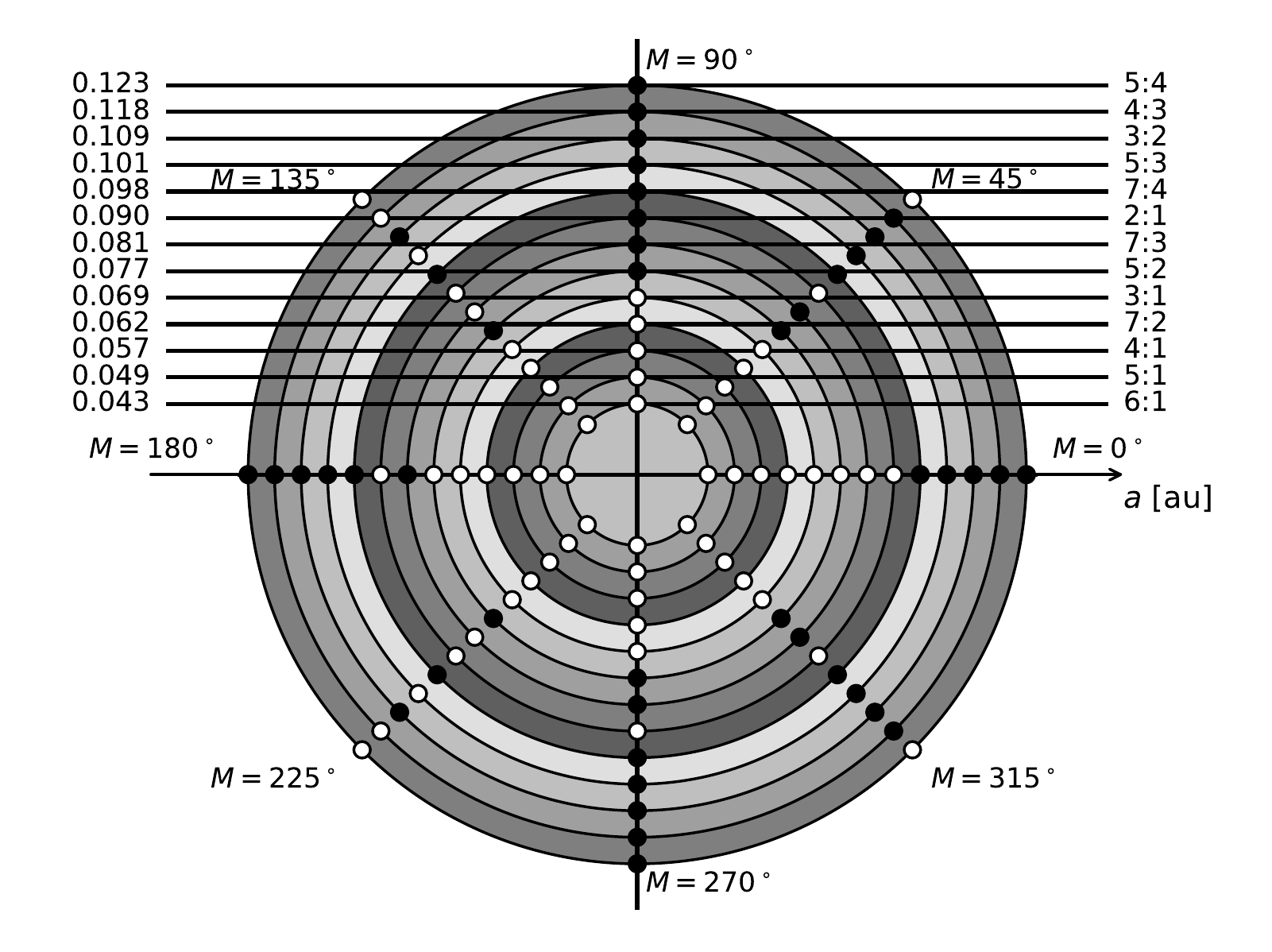}
\caption{Stability of resonances for different values of initial mean anomaly $M$. \textit{Top:} exterior resonances. \textit{Bottom:} interior resonances. White points indicate stable motion; black points show unstable motion.}
\label{fig:stab-res}
\end{figure}

Another possible explanation of periodic TTVs could be resonant perturbations between transiting planet Kepler-410Ab and another (unknown) planet with low mass (at the level of Earth mass) in this system. Resonant interactions are quite frequent among exoplanetary systems (e.g. \citealt{Fabrycky2012}). \cite{Wang2014} reported that mean-motion resonances (MMRs) 1:2 and 2:3 are most common in the systems observed by Kepler.

We used a method of upper mass limit \citep{Gibson2009} to determine the mass of a possible perturber. We set the amplitude of the observed TTVs as the amplitude of possible TTVs caused by a perturber. We assumed that the orbit of perturbing planet is circular and coplanar with that of transiting planet. We used Chamber's MERCURY6 code \citep{Chambers1999} to produce 20000 synthetic O-C diagrams for different configurations of mass and an orbital period of a hypothetical planet. We applied the Bulirsch-Stoer algorithm \citep{Stoer1982} to solve our three-body problem. Our calculation covered the whole length of the observing period ($\sim$ 1000 days). The results of our simulation are in Figure~\ref{fig:limit}. It displays the upper-mass limit of a hypothetical perturbing planet as a function of the ratio of the orbital period of perturbing planet ($P$) and that of the transiting planet Kepler-410Ab ($P_0$). Regions around $P/P_0 \approx 1 $ were found to be unstable due to frequent close encounters of the planets. Value of the upper mass limit is, in our case, the exact mass of possible planet orbiting with the period $P$. The existence of any planet above this limit is excluded by the amplitude of observed TTVs.

We could use the equation by \citet{Deck2016}:
\begin{equation}
\label{eq:P-OC-res}
P_{TTV} = \frac{1}{\left|j/P - (j-n)/P_0  \right|} 
\end{equation}
to calculate the the period of the perturbing planet where $P_{TTV}$ is a period of TTVs. We assume exterior MMRs $P_0 : P \approx j-n : j$ where $j$ and $n$ are small integer numbers and \mbox{$j > n \geq 1$}. Using this equation and the results shown in Figure \ref{fig:limit}, we could determine the mass and period of perturbing planet which generates TTV signal with a shape similar to the observed one displayed in Figure \ref{fig:ttv}. It is possible to find the solution for all kinds of resonances and we do not have any information about which resonance is correct. 

Following \cite{Freistetter2009}, we studied the stability of planets inside MMRs with Kepler-410Ab. In this simulation, we placed massless test particles directly inside a resonance. We performed this calculation for different interior ($P < P_0$) and exterior ($P > P_0$) resonances and different values of initial mean anomaly $M$. After $10^5$ revolutions of Kepler-410Ab, we analysed the stability of all resonances on a long-time scale.

Our results of stability of MMRs are shown in Figure \ref{fig:stab-res}. The exterior resonances are in general more stable than the interior. Only interior resonances with a higher resonance ratio are stable. Comparing the statistical representation of each resonance among the other exoplanetary systems \citep{Wang2014} and our stability study, the exterior resonance 2:3 with Kepler-410Ab looks the most probable. The planet with mass $\sim 1.5$\,M$_{Mars}$ and orbital period $\sim$\,26.5 days (close to the resonance 2:3) generates the TTV signal with a shape similar to the observed one. This planet would also be transiting (for coplanar orbits) but the depth of these transits (at a level of 0.01 mmag) is difficult to observe with current ground-based telescopes. We found no sign of transits of an additional planet in the Kepler-410 system in data from Kepler mission. The precision of Kepler is probably not sufficient to observe this very shallow transit. We believe that supposed precision of space missions Plato and CHEOPS \citep[about 10 ppm;][]{Marcos2014,Fortier2014} would allow detecting also such small exoplanets. 

A hypothetical planet with a higher resonance ratio to Kepler-410Ab would be larger and produce a deeper transit. However the probability of such resonance is smaller and so is the probability of observable transit. A quite common MMR 1:2 seems to be unstable in the system Kepler-410.

\section{Long-term stability}
\label{stab}

\begin{figure*}
\includegraphics[width=\textwidth]{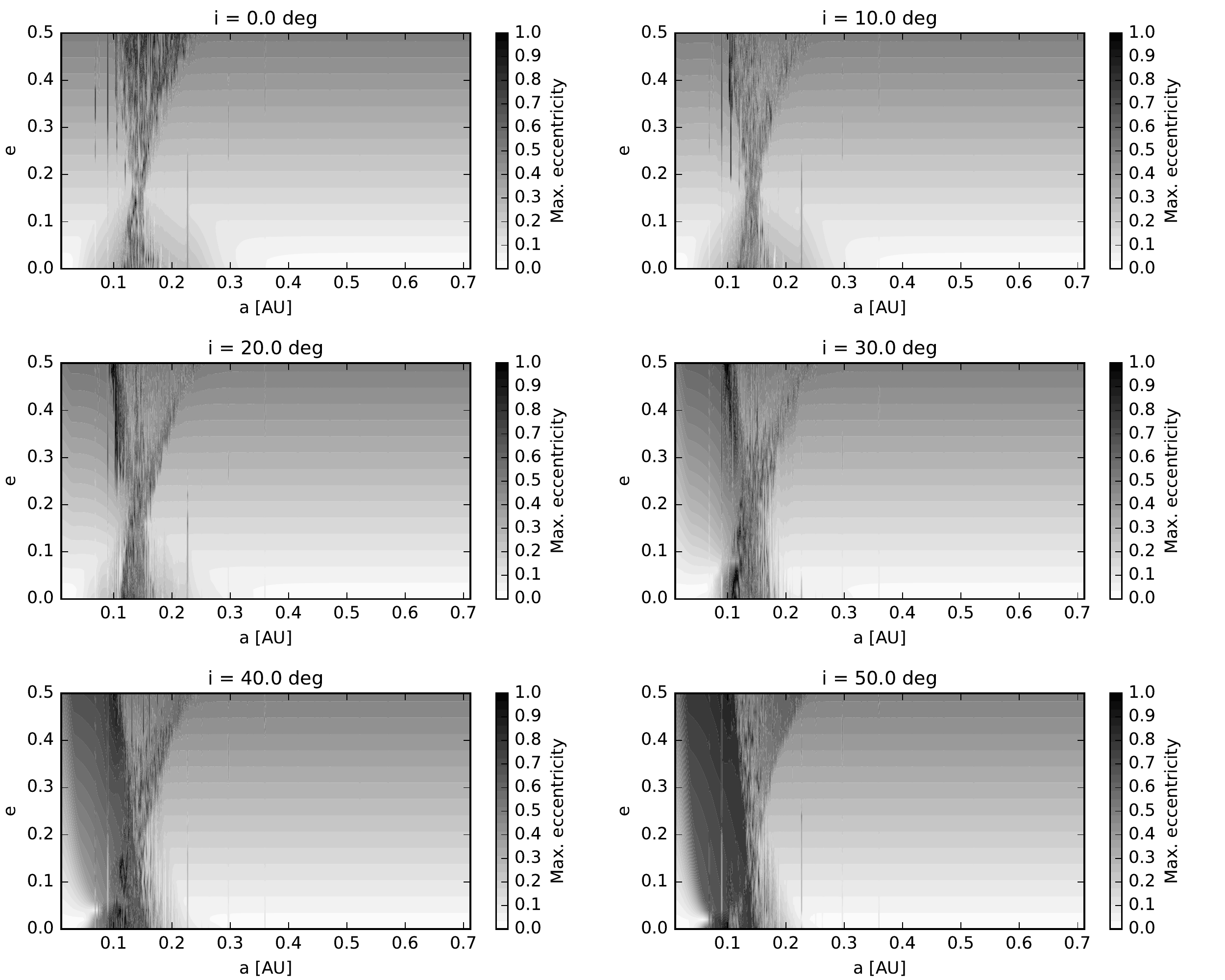}
\caption{Stability plot in the $a - e$ plane for different values of $i$ showing the maximum eccentricity (as an indicator of the stability of the orbit) for the system Kepler-410. From \textit{top left} to \textit{bottom right}: $i = 0\degr,\ i = 10\degr,\ i = 20\degr,\ i = 30\degr,\ i = 40\degr,\ i = 50\degr$.}
\label{fig:stab-ae}
\end{figure*}

\begin{figure*}
\includegraphics[width=\textwidth]{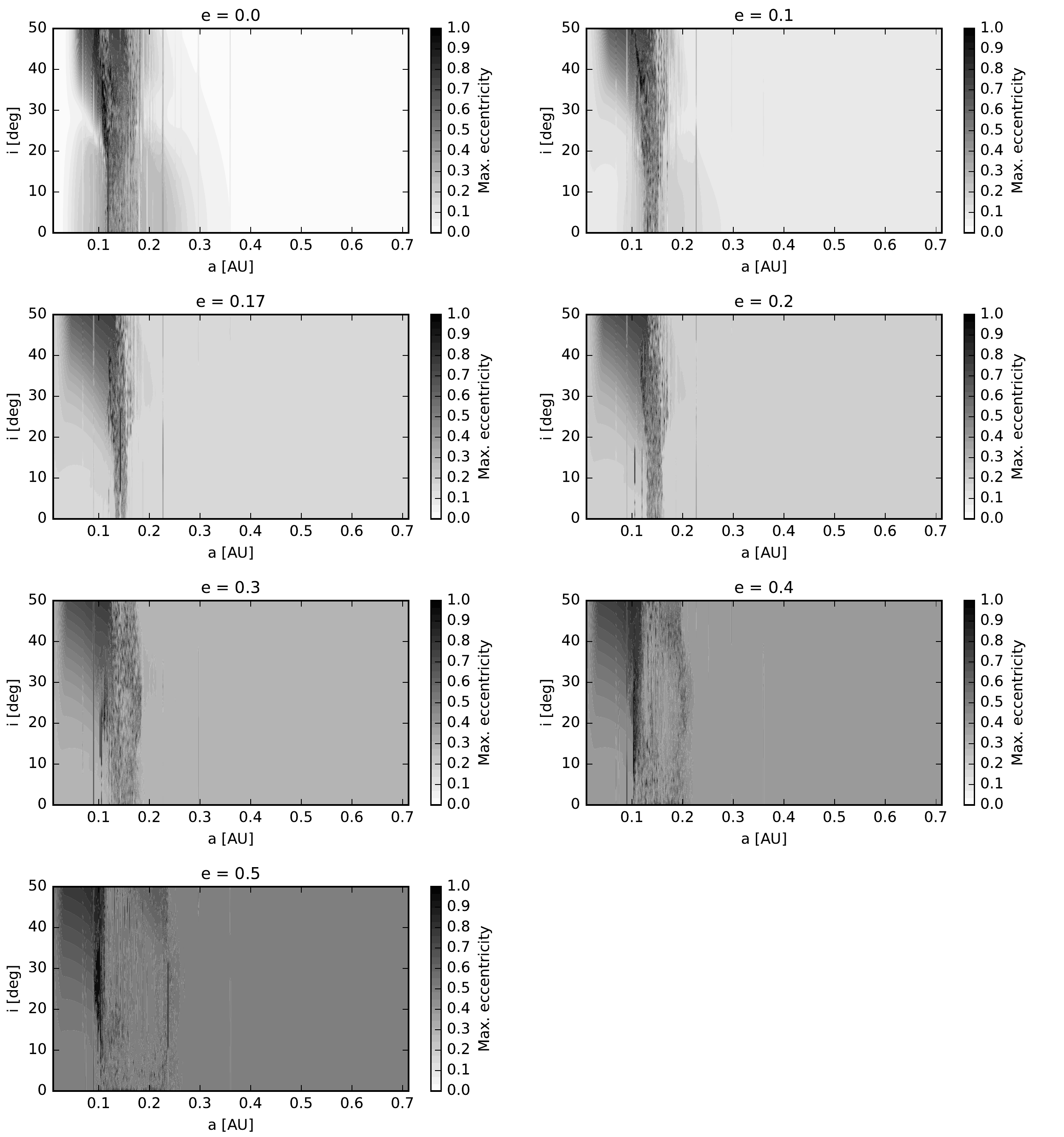}
\caption{Stability plot in the $a - i$ plane for different values of $e$ showing the maximum eccentricity for the system. From \textit{top left} to \textit{bottom left}: $e = 0.0,\ e = 0.1,\ e = 0.17,\ e = 0.2,\ e = 0.3,\ e = 0.4,\ e = 0.5$.}
\label{fig:stab-ai}
\end{figure*}

\begin{figure*}
\includegraphics[width=\textwidth]{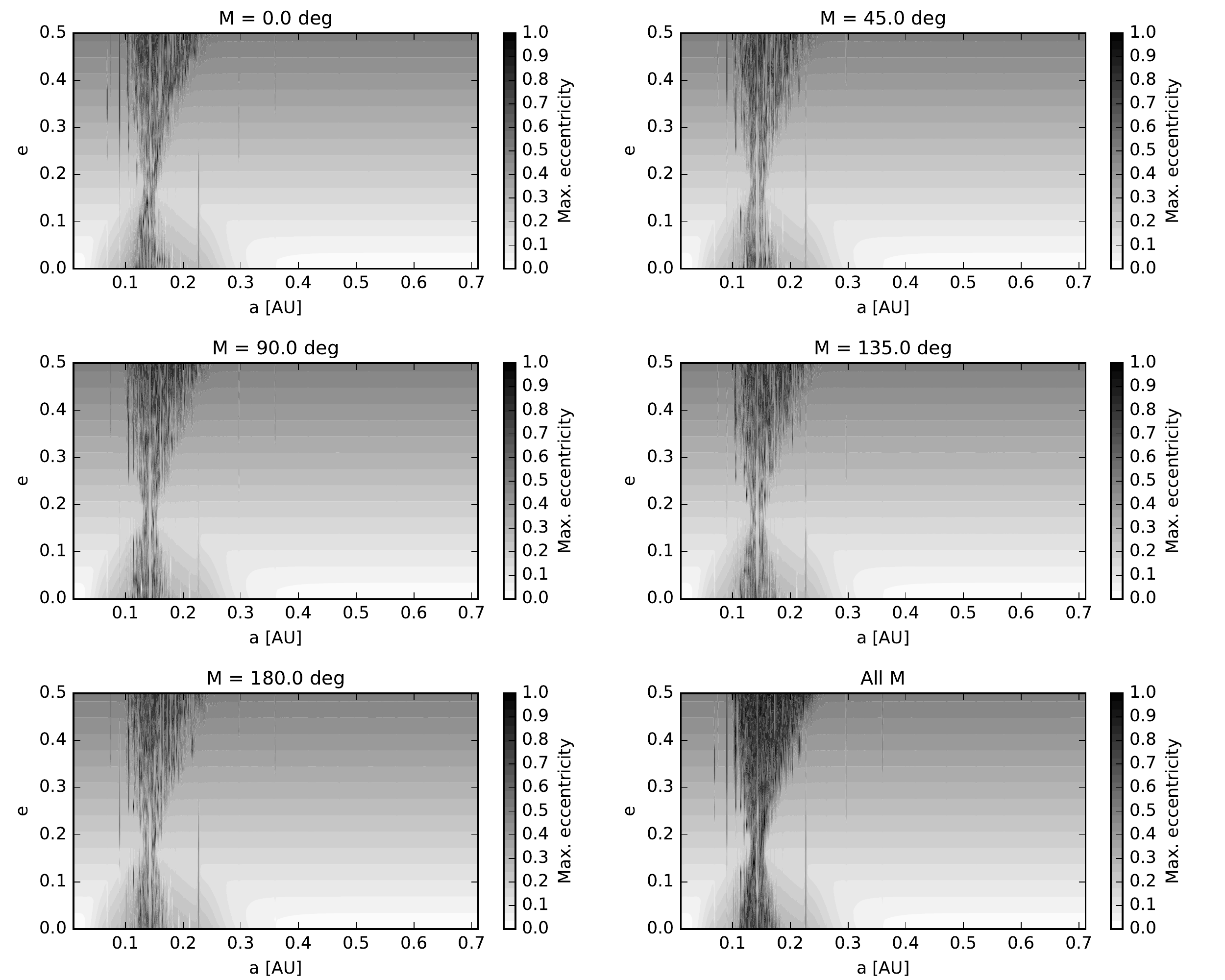}
\caption{Stability plot in the $a - e$ plane for different values of $M$ showing the maximum eccentricity for the system. From \textit{top left} to \textit{bottom left}: $M = 0\degr,\ M = 45\degr,\ M = 90\degr,\ M = 135\degr,\ M = 180\degr$. \textit{Bottom right}: combination of all figures using the maximal value of maximum eccentricity over all values of $M$.}
\label{fig:stab-M}
\end{figure*}

In this section, we study the long-term gravitational influence of the transiting planet on another potential planet in the system. We performed numerical simulations of the system with the parent star Kepler-410A, the transiting planet Kepler-410Ab and a large number of massless particles representing the hypothetical lower mass (earth mass or smaller) planets. The dwarf star Kepler-410B is too far to influence the stability of the planets on studied orbits.

We have used the method of the maximum eccentricity by \cite{Dvorak2003} to generate the stability maps. The value of maximum eccentricity of tested particle gives us information about the probability of a close encounter between the studied body and a massive planet Kepler-410Ab. An orbit with many close encounters will be, of course, unstable in the long term. Other authors (e.g. \citealt{Freistetter2009}) have noted that the maximum eccentricity is a good indicator of the stability of the orbit.

We performed numerical integration of the orbits for $10^5$ revolutions of Kepler-410Ab, giving a time span of approximately 4500 years. We again used the Bulirsch-Stoer algorithm and MERCURY6 code. The parameter $\epsilon$ which controls the accuracy of the integration was set to $10^{-8}$. 

We adopted the parameters of Kepler-410Ab from \cite{Eylen2014} for our simulations. We assumed eccentric orbit of Kepler-410Ab with eccentricity 0.17 \citep{Eylen2014}. The mass of Kepler-410Ab is unknown. We estimated its mass using the known radius and the mass-radius relation \mbox{$R \sim M^{0.55 \pm 0.02}$} \citep[up to $\sim$ 100~M$_{\oplus}$;][]{Bashi2017}. The calculated mass of Kepler-410Ab is $\sim$ 6~M$_{\oplus}$. To determine an accurate mass of Kepler-410Ab, one would need RV measurements with precision 1~m/s.

We have generated $3 \times 10^5$ massless particles with semi-major axes ranging from 0.01~au to 0.71~au (5 times the semi-major axis of Kepler-410Ab). The inner border was less than 2 times the star radius. The step for the semi-major axis was set to 0.0014~au (one hundredth of the semi-major axis of Kepler-410Ab). The upper limit of eccentricity was 0.5 and the step was 0.02. The mutual inclination of orbits of a test particle and the orbit of Kepler-410Ab was investigated from 0\degr to 50\degr with the step of 1\degr. Other orbital parameters (e.g. mean anomaly) were fixed to 0 in this simulation.

Figure \ref{fig:stab-ae} shows the stability maps in the $a - e$ plane for Kepler-410A system for selected values of inclination: $i = 0\degr,\ i = 10\degr,\ i = 20\degr,\ i = 30\degr,\ i = 40\degr,\ i = 50\degr$. The initial inclination of the orbit has minimal influence on the shape of the stability map in the $a - e$ plane. It plays some role in the case of orbits closer to the parent star. Inner orbits are also stable up to inclinations of 30\degr. On the other hand, the impact of the initial eccentricity on the stability of the orbit is important. The shape of unstable regions is similar to a double cone with the apex at the assumed eccentricity ($e = 0.17$) of Kepler-410Ab. The other objects with a different value of eccentricity cross the orbit of Kepler-410Ab also on orbits with the semi-major axis which differ more from that of Kepler-410Ab. And it causes a higher probability of a close encounter with the giant planet. The biggest number of possible orbits is therefore for eccentricities $e \approx 0.17$.

In Figure \ref{fig:stab-ai}, we present stability maps in the $a - i$ plane for the system for selected initial values of eccentricity: $e = 0.0,\ e = 0.1,\ e = 0.17,\ e = 0.2,\ e = 0.3,\ e = 0.4,\ e = 0.5$. The width of the stable regions is generally almost independent of the inclination, mainly for lower values of eccentricity. For higher values of eccentricity, there is a wider unstable region for the lower inclinations. 

We also studied effect of the initial mean anomaly on stability. We generated the stability maps in the $a-e$ plane for selected values of the mean anomaly: $M = 0\degr,\ M = 45\degr,\ M = 90\degr,\ M = 135\degr,\ M = 180\degr$. We were interested only in coplanar orbits (i.e.  $i = 0\degr$). Results of this simulation are shown in Figure \ref{fig:stab-M}. The last picture on this figure was created by taking the global maximum eccentricity over all values of the initial mean anomaly. In all cases, the shape of the stable regions looks the same, i.e. it is not influenced by the mean anomaly. However, additional stable and unstable regions corresponding with the MMRs are strongly dependent on the initial mean anomaly.

\section{Conclusions}
\label{concl}

We performed spectroscopic observations of Kepler-410 during three years. In all obtained spectra, we detected spectral lines only from one source (Kepler-410A). We made 60 RV measurements. We did not observe any periodic variation of the radial velocities with amplitude more than 800~m/s over three observing seasons. Therefore, we rejected our previous hypothesis that observed TTVs are caused by the presence of another stellar companion in the system \citep{Gajdos2017}. We can also exclude the existence of a brown dwarf (or any other more massive object) down to a mass of about 30 $M_{Jup}$ on the orbits with period up to 2 years or a massive hot Jupiter in the Kepler-410 system based on the accuracy of our RV measurements on nearly edge-on orbits.

Another logical explanation of the observed TTVs is gravitational interaction with an additional planet orbiting near to the MMR. Our analysis of the stability of the most common interior and exterior resonances shows that the exterior MMR 2:3 is the most probable. Small planet ($\sim 1.5$~M$_{Mars}$) on an orbit with period $\sim$~26.5 days (close to resonance 2:3) would cause a similar TTVs as observed in Kepler-410Ab. Unfortunately, this planet would be so small that its observational evidence is hardly possible to get at present time. This additional planet would be transiting with depth of transit at the level of only 0.01 mmag. The radial-velocity measurements with precision 1 cm/s would be required to detect this planet spectroscopically. The successful observation would be also complicated by the stellar oscillations of Kepler-410A \citep{Eylen2014}. However, prepared space missions Plato and CHEOPS should detect also similarly small exoplanets.

Finally, we investigated the dynamical stability of the system in order to identify stable regions where additional planets could exist for a long time (hundreds of years). We found that a hypothetical planet could exist relatively close to the known transiting planet, mainly on coplanar orbit with eccentricity similar to that of known planet Kepler-410Ab ($e=0.17$). Raising or reduction of the eccentricity from this value causes the unstable region to start to grow. Orbits with semi-major axis more than $\sim 0.25$~au are stable for almost all combinations of eccentricities and inclinations (except some MMRs). The inner orbits are stable for orbital inclinations up to 20\degr -- 30\degr. 

From a dynamic point of view, the possibility of other planets in the system is high. There exists a wide range of orbits on which such planets could exist for a long time.

\section*{Acknowledgement}
We are grateful to Z. Garai, E. Kundra, E. Paunzen, R.\,Kom\v{z}\'ik and P. Sivani\v{c} for help in obtaining the observations of Kepler-410 at Observatory Star\'a Lesn\'a and Skalnat\'e Pleso. This paper has been supported by the grant of the Slovak Research and Development Agency with number APVV-15-0458 and was created under project ITMS No.26220120029, based on the supporting operational Research and development program financed from the European Regional Development Fund. The research of P.G. was supported by internal grant VVGS-PF-2017-724 of the Faculty of Science, P. J. \v{S}af\'{a}rik University in Ko\v{s}ice. M.V., T.P. and \v{L}.H. would like to thank the project VEGA 2/0031/18. P.K., J.S. and D.D. would like to acknowledge the support from GACR international grant 17-01752J. M.S. acknowledges financial support of Postdoc@MUNI project CZ$.02.2.69/0.0/0.0/16\_027/0008360$.

\label{lastpage}


\begin{thebibliography}{}
\makeatletter
\relax
\def\mn@urlcharsother{\let\do\@makeother \do\$\do\&\do\#\do\^\do\_\do\%\do\~}
\def\mn@doi{\begingroup\mn@urlcharsother \@ifnextchar [ {\mn@doi@}
  {\mn@doi@[]}}
\def\mn@doi@[#1]#2{\def\@tempa{#1}\ifx\@tempa\@empty \href
  {http://dx.doi.org/#2} {doi:#2}\else \href {http://dx.doi.org/#2} {#1}\fi
  \endgroup}
\def\mn@eprint#1#2{\mn@eprint@#1:#2::\@nil}
\def\mn@eprint@arXiv#1{\href {http://arxiv.org/abs/#1} {{\tt arXiv:#1}}}
\def\mn@eprint@dblp#1{\href {http://dblp.uni-trier.de/rec/bibtex/#1.xml}
  {dblp:#1}}
\def\mn@eprint@#1:#2:#3:#4\@nil{\def\@tempa {#1}\def\@tempb {#2}\def\@tempc
  {#3}\ifx \@tempc \@empty \let \@tempc \@tempb \let \@tempb \@tempa \fi \ifx
  \@tempb \@empty \def\@tempb {arXiv}\fi \@ifundefined
  {mn@eprint@\@tempb}{\@tempb:\@tempc}{\expandafter \expandafter \csname
  mn@eprint@\@tempb\endcsname \expandafter{\@tempc}}}

\bibitem[{Adams} et~al.(2012)]{Adams2012}
{Adams} E.~R.,  {Ciardi} D.~R.,  {Dupree} A.~K.,  {Gautier} T.~N. I.,  {Kulesa}
  C.,   {McCarthy} D.,  2012, \mn@doi [\aj] {10.1088/0004-6256/144/2/42}, \href
  {https://ui.adsabs.harvard.edu/#abs/2012AJ....144...42A} {144, 42}

\bibitem[{Agol} et~al.(2005)]{Agol2005}
{Agol} E.,  {Steffen} J.,  {Sari} R.,   {Clarkson} W.,  2005, \mn@doi [\mnras]
  {10.1111/j.1365-2966.2005.08922.x}, \href
  {https://ui.adsabs.harvard.edu/#abs/2005MNRAS.359..567A} {359, 567}

\bibitem[{Bashi} et~al.(2017)]{Bashi2017}
{Bashi} D.,  {Helled} R.,  {Zucker} S.,   {Mordasini} C.,  2017, \mn@doi [\aap]
  {10.1051/0004-6361/201629922}, \href
  {http://adsabs.harvard.edu/abs/2017A%26A...604A..83B} {604, A83}

\bibitem[{Baudrand} \&
  {Bohm}(1992)]{Baudrand1992}
{Baudrand} J.,  {Bohm} T.,  1992, \aap, \href
  {http://adsabs.harvard.edu/abs/1992A%26A...259..711B} {259, 711}

\bibitem[{Blanco-Cuaresma} et~al.(2014)]{Cuaresma2014}
{Blanco-Cuaresma} S.,  {Soubiran} C.,  {Heiter} U.,   {Jofr{\'e}} P.,  2014,
  \mn@doi [\aap] {10.1051/0004-6361/201423945}, \href
  {http://adsabs.harvard.edu/abs/2014A%26A...569A.111B} {569, A111}

\bibitem[{Chambers}(1999)]{Chambers1999}
{Chambers} J.~E.,  1999, \mn@doi [\mnras] {10.1046/j.1365-8711.1999.02379.x},
  \href {https://ui.adsabs.harvard.edu/#abs/1999MNRAS.304..793C} {304, 793}

\bibitem[{Deck} \&
  {Agol}(2016)]{Deck2016}
{Deck} K.~M.,  {Agol} E.,  2016, \mn@doi [\apj] {10.3847/0004-637X/821/2/96},
  \href {https://ui.adsabs.harvard.edu/#abs/2016ApJ...821...96D} {821, 96}

\bibitem[{Dvorak} et~al.(2003)]{Dvorak2003}
{Dvorak} R.,  {Pilat-Lohinger} E.,  {Funk} B.,   {Freistetter} F.,  2003,
  \mn@doi [\aap] {10.1051/0004-6361:20021805}, \href
  {https://ui.adsabs.harvard.edu/#abs/2003A&A...398L...1D} {398, L1}

\bibitem[{Fabrycky}
  et~al.(2012)]{Fabrycky2012}
{Fabrycky} D.~C.,  et~al., 2012, \mn@doi [\apj] {10.1088/0004-637X/750/2/114},
  \href {http://adsabs.harvard.edu/abs/2012ApJ...750..114F} {750, 114}

\bibitem[{Fortier} et~al.(2014)]{Fortier2014}
{Fortier} A.,  {Beck} T.,  {Benz} W.,  {Broeg} C.,  {Cessa} V.,  {Ehrenreich}
  D.,   {Thomas} N.,  2014, in Space Telescopes and Instrumentation 2014:
  Optical, Infrared, and Millimeter Wave. p. 91432J,
  \mn@doi{10.1117/12.2056687}

\bibitem[{Freistetter} et~al.(2009)]{Freistetter2009}
{Freistetter} F.,  {S{\"u}li} {\'A}.,   {Funk} B.,  2009, \mn@doi [Astron.
  Nachr.] {10.1002/asna.200811199}, \href
  {https://ui.adsabs.harvard.edu/#abs/2009AN....330..469F} {330, 469}

\bibitem[{Gaia
  Collaboration} et~al.(2018)]{Gaia2018}
{Gaia Collaboration} et~al., 2018, \mn@doi [\aap]
  {10.1051/0004-6361/201833051}, \href
  {https://ui.adsabs.harvard.edu/#abs/2018A&A...616A...1G} {616, A1}

\bibitem[{Gajdo{\v{s}}} et~al.(2017)]{Gajdos2017}
{Gajdo{\v{s}}} P.,  {Parimucha} {\v{S}}.,  {Hamb{\'a}lek} {\v{L}}.,
  {Va{\v{n}}ko} M.,  2017, \mn@doi [\mnras] {10.1093/mnras/stx963}, \href
  {https://ui.adsabs.harvard.edu/#abs/2017MNRAS.469.2907G} {469, 2907}

\bibitem[{Gibson}
  et~al.(2009)]{Gibson2009}
{Gibson} N.~P.,  et~al., 2009, \mn@doi [\apj] {10.1088/0004-637X/700/2/1078},
  \href {https://ui.adsabs.harvard.edu/#abs/2009ApJ...700.1078G} {700, 1078}

\bibitem[{Holczer}
  et~al.(2016)]{Holczer2016}
{Holczer} T.,  et~al., 2016, \mn@doi [\apjs] {10.3847/0067-0049/225/1/9}, \href
  {http://adsabs.harvard.edu/abs/2016ApJS..225....9H} {225, 9}

\bibitem[{Irwin}(1952a)]{Irwin1952}
{Irwin} J.~B.,  1952a, \mn@doi [\apj] {10.1086/145604}, \href
  {https://ui.adsabs.harvard.edu/#abs/1952ApJ...116..211I} {116, 211}

\bibitem[{Irwin}(1952b)]{Irwin1952-rv}
{Irwin} J.~B.,  1952b, \mn@doi [\apj] {10.1086/145605}, \href
  {https://ui.adsabs.harvard.edu/#abs/1952ApJ...116..218I} {116, 218}

\bibitem[{Kabath}
  et~al.(2018)]{Kabath2018}
{Kabath} P.,  {Skarka} M.,   {Sabotta} S.,  2018, submitted to \aj

\bibitem[{Marcos-Arenal}
  et~al.(2014)]{Marcos2014}
{Marcos-Arenal} P.,  et~al., 2014, \mn@doi [\aap]
  {10.1051/0004-6361/201323304}, \href
  {https://ui.adsabs.harvard.edu/\#abs/2014A&A...566A..92M} {566, A92}

\bibitem[{Mazeh}
  et~al.(2013)]{Mazeh2013}
{Mazeh} T.,  et~al., 2013, \mn@doi [\apjs] {10.1088/0067-0049/208/2/16}, \href
  {https://ui.adsabs.harvard.edu/#abs/2013ApJS..208...16M} {208, 16}

\bibitem[{Molenda-{\.Z}akowicz} et~al.(2013)]{Molenda2013}
{Molenda-{\.Z}akowicz} J.,  et~al., 2013, \mn@doi [\mnras]
  {10.1093/mnras/stt1095}, \href
  {https://ui.adsabs.harvard.edu/#abs/2013MNRAS.434.1422M} {434, 1422}

\bibitem[{Nidever} et~al.(2002)]{Nidever2002}
{Nidever} D.~L.,  {Marcy} G.~W.,  {Butler} R.~P.,  {Fischer} D.~A.,   {Vogt}
  S.~S.,  2002, \mn@doi [\apjs] {10.1086/340570}, \href
  {https://ui.adsabs.harvard.edu/#abs/2002ApJS..141..503N} {141, 503}

\bibitem[{Pribulla}
  et~al.(2015)]{Pribulla2015}
{Pribulla} T.,  et~al., 2015, \mn@doi [Astron. Nachr.]
  {10.1002/asna.201512202}, \href
  {https://ui.adsabs.harvard.edu/#abs/2015AN....336..682P} {336, 682}

\bibitem[Stoer \&
  Bulirsch(1980)]{Stoer1982}
Stoer J.,  Bulirsch R.,  1980, Introduction to numerical analysis.
Springer-Verlag, New York

\bibitem[{Tody}(1993)]{Tody1993}
{Tody} D.,  1993, in {Hanisch} R.~J.,  {Brissenden} R.~J.~V.,   {Barnes} J.,
  eds,  Astronomical Society of the Pacific Conference Series Vol. 52,
  Astronomical Data Analysis Software and Systems II. p.~173

\bibitem[{Van Eylen}
  et~al.(2014)]{Eylen2014}
{Van Eylen} V.,  et~al., 2014, \mn@doi [\apj] {10.1088/0004-637X/782/1/14},
  \href {https://ui.adsabs.harvard.edu/#abs/2014ApJ...782...14V} {782, 14}

\bibitem[{Wang} \&
  {Ji}(2014)]{Wang2014}
{Wang} S.,  {Ji} J.,  2014, \mn@doi [\apj] {10.1088/0004-637X/795/1/85}, \href
  {https://ui.adsabs.harvard.edu/#abs/2014ApJ...795...85W} {795, 85}

\makeatother
\end{thebibliography}
\end{document}